\documentclass[conference]{IEEEtran}
\usepackage[utf8]{inputenc}
\usepackage{graphicx}
\usepackage{epstopdf}
\usepackage{amsfonts}
\usepackage{amsmath}
\usepackage{color}
\usepackage{cite}
\usepackage{mathtools}
\newcommand\norm[1]{\left\lVert#1\right\rVert}
\usepackage{stmaryrd}
\usepackage{algpseudocode}% http://ctan.org/pkg/algorithmicx
\usepackage[linesnumbered,ruled,vlined]{algorithm2e}
\usepackage{url}
\usepackage[thinlines]{easytable}
\usepackage{makecell}

\usepackage[linesnumbered,ruled]{algorithm2e}

\newcommand*\diag {\mathop{}\!\mathrm{diag}}

\DeclareMathOperator*{\argmaxA}{arg\,max} % Jan Hlavacek
\DeclareMathOperator*{\argminA}{arg\,min} % Jan Hlavacek
\ifCLASSINFOpdf

\else
\fi
\begin{document}
\title{Sequence Block based  Compressed Sensing Multiuser Detection for 5G}
\author{\IEEEauthorblockN{Mehmood Alam,  Qi Zhang}
\IEEEauthorblockA{Department of Engineering\\ Aarhus University, Denmark
\\Email:\{mehmood.alam, qz\}@eng.au.dk
}}
\maketitle
\begin{abstract}
Compressed sensing based multiuser detection (CSMUD) is a promising candidate to cope with the massive connectivity requirements of the massive machine type communication (mMTC) in the fifth generation (5G) wireless communication system. It facilitates grant-free non-orthogonal code division multiple access (CDMA) to accommodate massive number of IoT devices. In non-orthogonal CDMA, the users are assigned with  non-orthogonal sequences which serve as their signatures.  However, the activity detection which is based on the correlation between the spreading sequences, degrades with increase in the number of users, especially in the lower SNR region. In this paper, to improve the performance of the CSMUD, we propose a sequence block based CSMUD, in which block of sequences is used as signature of the user instead of single sequence. A sequence block based group orthogonal matching pursuit algorithm is proposed to jointly detect the activity and data. The proposed scheme reduces the  detection error rate (DER)  by a magnitude of two at SNR = 10 dB in a system where the number of users are three times more than the number of available resources with each user having activity probability of 0.1. The DER of the proposed scheme is below $10^{-2}$ even at activity probability of 0.16 for sequence length of 20 and  overloading factor of 300\%. Furthermore, at SNR = 10 dB, the DER of the proposed scheme outperforms the conventional scheme by a magnitude of one for a system with overloading factor of 500\%.

 \end{abstract}
\IEEEpeerreviewmaketitle
\section{Introduction}  
   In modern wireless communication system, the number of devices is dramatically increasing and it is estimated that the number of connected devices will  rise to 50 billion  by 2020 \cite{evans2011internet} and even up to 500 billion by 2024 \cite{500}. The dominant percentage of these devices will be IoT devices which are characterized by sporadic activity and low data rate. The substantial increase in the number of IoT devices imposes enormous challenges  to meet the performance requirements such as high spectral efficiency, massive connectivity,  low power consumption, ultra-high reliability and ultra-low latency of the fifth generation (5G) communication system \cite{zhang2015mission} \cite{wp}. The  current orthogonal multiple access (OMA) schemes are based on orthogonal resource allocation according to which  only  one user is served in a resource block, e.g., a single time slot, a spreading code or  a frequency channel. Due to the orthogonality constraint, i.e., the number of users cannot exceeds the orthogonal resources; the OMA schemes are not capable to meet the performance requirements of the 5G communication  system. \par   
   Non-orthogonal multiple access (NOMA) has become a key technique to meet the performance requirements of the 5G communication system. In NOMA schemes, unlike the conventional orthogonal multiple access (OMA) schemes, non-orthogonal resources are assigned to the users. The non-orthogonality allows the multiple access scheme to be overloaded, i.e., to serve multiple users in each orthogonal resource block.  Various schemes have been proposed in the literature that uses the NOMA principal, which can be divided into two main categories: the power domain NOMA and the code domain NOMA. In the power domain NOMA, the non-orthogonality is introduced by allocating different power levels to the users \cite{pnoma}. The power levels are unique for each user and serve as signatures in the multiuser detection (MUD). The MUD is carried out by using successive interference cancellation (SIC) \cite{sic}. \par 
 In code domain NOMA, the users are allowed to transmit their data by using non-orthogonal codes. At the receiver, advanced MUD techniques such as message passing algorithm \cite{mpa} and SIC are used to separate the users. There are different variants of the code domain NOMA, the most promising of which are sparse code multiple access (SCMA) \cite{scma, MC, 7925767}, pattern division multiple access (PDMA) \cite{pdma}, multi-user shared access (MUSA) \cite{musa} and so on. In SCMA, the information bits after channel coding are directly mapped to user specific sparse codewords and message passing algorithm  is used at the receiver for MUD. The PDMA considers each user’s channel state and accordingly allocates a different number of non-zero elements to the codeword. The MUSA is a sequence based NOMA scheme that assigns low correlated spreading sequences to the nodes and uses SIC for multiuser detection. \par
Compressive sensing based multiuser detection (CSMUD) \cite{csp1, 6125356, MCSM, DBLP, ECSMUD} is a recently proposed scheme to enable grant-free non-orthogonal code division multiple access (CDMA) for sporadic  mMTC in 5G. Non-orthogonal spreading sequences are assigned to the users for spreading their data. At the receiver, the CSMUD exploits the sporadic activity of the IoT devices to jointly detect the activity and data using compressive sensing algorithms, e.g., orthogonal matching pursuit (OMP) \cite{omp1}. OMP is the preferred choice for CSMUD due to its relative low complexity. A variant of the OMP called group orthogonal matching pursuit (GOMP) is used in \cite{MCSM}. The GOMP exploits the group sparsity, i.e., when a user is active, it transmits a block of symbols. However, the performance of the  OMP based CSMUD algorithms depends on the correlation between the spreading sequences: higher the correlation  is, higher the detection error rate will be. For fixed number of sequences, the correlation between the sequences is dependent on the spreading factor, i.e., the length of the spreading sequences.  However, it is spectrally inefficient to  reduce the mutual correlation of the sequences by increasing the spreading factor.\par   
In this paper, we have proposed a sequence block based CSMUD (SB-CSMUD) to improve the activity detection in CSMUD without increasing the spreading factor. In the proposed scheme, a block of sequences is assigned to each user instead of assigning a single sequence. The activity detection is based on the correlation of the sequence blocks instead of the correlation between single sequences. The blocks are designed using the total available spreading sequences.  A sequence block GOMP (SB-GOMP) algorithm is proposed to jointly detect the activity and data. The proposed algorithm uses the sequence blocks as signature to distinguish the users.  Due to averaging the correlation of the sequences in a block, the maximum correlation between the blocks of sequences is less than the maximum correlation between single sequences, which significantly improves the activity detection. It is shown that for a block size of 4, without increasing the spreading factor, the DER is reduced by magnitude of two at SNR = 10 dB. Furthermore, the proposed scheme is resilient at higher activity probability, i.e., when more  devices are active at a time  and when the number of devices is higher  in the range of base station, the network has  higher overloading. It is shown that at SNR = 10 dB, a gain of magnitude one in DER is achieved over the conventional scheme at overloading factor of five for spreading factor of 20 and sequence block size of four.  In addition, a gain of magnitude one is achieved over the conventional scheme at activity probability of 0.16. \par
The paper is organized as follows. In Section \ref{sysmod}, the general system model for mMTC and compressive sensing basics are described. In Section \ref{proposed}, the proposed method of sequence block based compressive sensing multiuser detection is described. In Section \ref{pa}, the simulation parameters are given and the performance of the proposed scheme is analyzed in terms of detection error rate and bit error rate. Finally Section \ref{con} concludes the paper.\par
\textit{Notations:} In this paper, all boldface uppercase letters represent matrices such as $\mathbf{A}$, while all lowercase boldface letters represent vectors such as $\mathbf{x}$. The set of binary and complex numbers are represented by $\mathbb{B}$ and $\mathbb{C}$, respectively.  Italic letters such as $k$, $x$ represent variables. Uppercase letters such as $K$ and Greek letters such as $\gamma$ represent a constant value.  
\section{System Model} \label{sysmod}
\begin{figure}[!h]
	\centering
	\includegraphics[scale=.75]{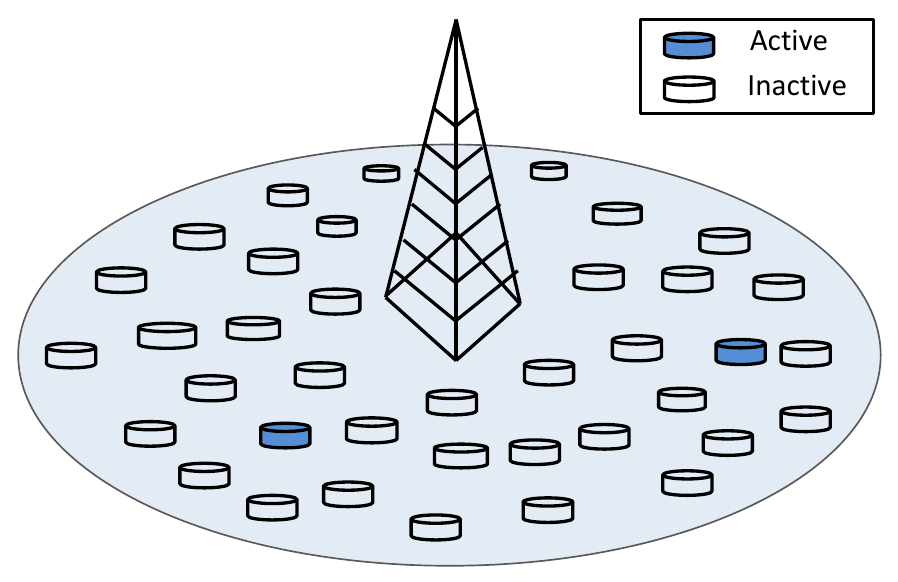}
	\caption{Massive machine type communication uplink scenario: $N$ nodes connected to the base station}
	\label{sc}
\end{figure}

A typical uplink mMTC scenario is considered where  $N$ users are in the range of a base station as shown in Figure \ref{sc}.  It is assumed that each user is active with an activity probability,  $p_a\ll 1$. With such a low activity probability, the data transmission is sporadic, i.e., in a single time slot only a small fraction of the $N$ users is active.  
 Furthermore, we assume that when a node is active, it transmits $L$ consecutive bits. Under these assumptions, we modeled the mMTC traffic as a Bernoulli distribution with activity probability, $p_a\ll 1$, and inactive probability $1-p_a$. The low activity probability  assumption makes the multiuser signal sparse, i.e., a small number of non-zero elements. The multiuser detection problem for such sparse signals can be modeled as a compressed sensing problem.\par
Compressed sensing (CS) is a signal processing technique which samples a sparse signal at a rate much less than the Nyquist rate \cite{cs}. A signal  $\mathbf{v}\in\mathbb{C}^{N\times1}$ is said to be $K$-sparse if it has only $K$ non-zero elements. The CS process produces measurement $\mathbf{y} \in \mathbb{C}^{M\times 1}$ by a measurement matrix  $\mathbf{\Psi}\in \mathbb{C}^{M\times N}$, $K<M<N$,
\begin{equation}
\mathbf{y}=\mathbf{\Psi}\mathbf{v}.
\label{cseq}
\end{equation}
As $M<N$, Equation (\ref{cseq}) is an underdetermined system of equations and  convex optimization   is used to reconstruct the signal $\mathbf{v}$ as follows 
\begin{equation}
\hat{\mathbf{v}}= \argminA_{\mathbf{v} \in \mathbb{C}^{N\times1}} {\Vert \mathbf{v} \Vert}_1   \:\:\:  \text{subject to   }\: \:  \mathbf{y} =\mathbf{\Psi} \mathbf{v}.
\end{equation}
where ${\Vert \mathbf{v} \Vert}_1 $ represents the $l_1$ norm of $\mathbf{v}$. Convex optimization recovers the signal with higher accuracy, however, it has a complexity of cubic order. Due to the higher complexity, the use of convex optimization becomes impractical when the number of $N$ is higher. Therefore, greedy algorithms such as OMP and orthogonal least square (OLS) are used for signal recovery with relatively lower computational complexity \cite{cs}. In OMP the support of $\mathbf{v}$ is iteratively obtained by selecting the index of the maximally correlated column of $\mathbf{\Psi}$  with the residual, whereas in OLS a least square criteria is used instead of correlation. The residual is initialized as $\mathbf{y}$ and is updated in each iteration of the greedy algorithm. The data of the detected active users are then estimated using least square estimation. However, for the successful recovery of signal using CS algorithms, the measurement matrix, $\mathbf{\Psi}$, needs to satisfy the restricted isometry property (RIP) \cite{candes2005decoding}. A matrix  $\mathbf{\Psi}$ is said to satisfy RIP if there exists a constant $\delta_{K} \in (0,1)$ such that for every $K$ sparse signal, $\mathbf{v}$,  \cite{candes2005decoding}
\begin{equation}
(1-\delta_{K})\norm{\mathbf{v}}_2^2\leq \norm{ \mathbf{\Psi}\mathbf{v}}_2^2\leq (1+\delta_{K})\norm{\mathbf{v}}_2^2.
\end{equation}\par  
In the proposed work we assume  the non-orthogonal multicarrier-CDMA (MC-CDMA) system \cite{mccdma} for uplink mMTC scenario. For the MUD problem in our system model, the column vector $\mathbf{v}$ in Equation (\ref{cseq})  can be regarded as the multiuser signal which represents the data symbols of $N$ users in a single time slot. $M$ represents the number of radio resources  and   $K$ is the number of simultaneously active users which is dependent on the activity probability, $p_a$.  Generally, the sensing matrix, $\mathbf{\Psi}$,  consists of the combination of the spreading sequences  and channel coefficients. However, in the proposed scheme, differential quadrature phase shift keying (DQPSK) is used  to avoid the channel estimation, therefore, the  sensing matrix used for recovering the signal only consists of the spreading sequences.
\section{Sequence block based CSMUD} \label{proposed}
 In conventional CSMUD based NOMA scheme, each user is assigned with a specific spreading sequence. A node uses the same sequence for all its data symbols in a frame.  For sensing matrix, $\mathbf{A}=[\mathbf{a}_1 \: \:\:\: \mathbf{a}_2  \hdots \mathbf{a}_n  \hdots   \mathbf{a}_N]$, where $\mathbf{a}_n$ is spreading sequence assigned to user $n$, the GOMP based activity detection at iteration $i$ is given as
\begin{equation}
\label{ac1}
I= \argmaxA_{1\leq j \leq N} \left(\Omega_j- \langle \mathbf{a}_{_j}, \mathbf{a}_{(k)}\rangle \right)
\end{equation}
where $\mathbf{a}_{(k)}$ is the detected sequence in the previous iteration and $\Omega_j$, $1\le j \leq N$,  is given as  

\begin{equation}
\label{c1}
\Omega_j= \frac{1}{L} \sum\limits_{\substack{l=1}}^{L}\sum\limits_{i=1}^{N} {x}_i^{(l)} \cdot \langle\mathbf{a}_{_j},\mathbf{a}_{_i}\rangle+w_j^{(l)},
\end{equation}
where $w_j^{(l)}$ is the contribution of noise at $l$-th symbol of user $j$. From Equation (\ref{c1}), it is evident that activity is dependent on the correlation between the sequences and a detection error occurs when the sum of the cross correlations  of an inactive node becomes higher than that of the active node. Therefore, the performance of the GOMP based CSMUD depends on the maximum correlation, $\mu$,  between the sequences of a  sensing matrix, $\mathbf{A}$, which is defined as 
\begin{equation}
\label{mu}
\mu= \max\limits_{\tiny{\substack{i,j\\ i\neq j}}} \frac{|\mathbf{a}_i^T\mathbf{a}_j|}{\|\mathbf{a}_i\|\|\mathbf{a}_j\|} \:\:\:\:\:   i,j \in \{1,2 \hdots N \} 
\end{equation}\par 
For a sensing matrix  with higher value of $\mu$,  the  detection error rate (DER) will be higher and vice versa. The maximum correlation, $\mu$,  is  dependent on the length of the spreading sequence, i.e, the spreading factor. Increasing the spreading factor reduces the maximum correlation $\mu$, which improves the activity detection. However, higher spreading factor means using more radio resources, which is not desirable in a mMTC scenario. Therefore, to reduce the $\mu$ value without increasing the spreading factor, we propose SB-CSMUD. In SB-CSMUD, instead of assigning a single sequence, a sequence block, $\mathbf{B}_n \subset \mathbf{A}, 1 \leq n \leq N $, is  assigned to each user and accordingly a  sequence block based GOMP algorithm is designed for multiuser detection. 
\subsection{Sequence Block Design}
\begin{figure}[!h]
	\centering
	\includegraphics[scale=.4]{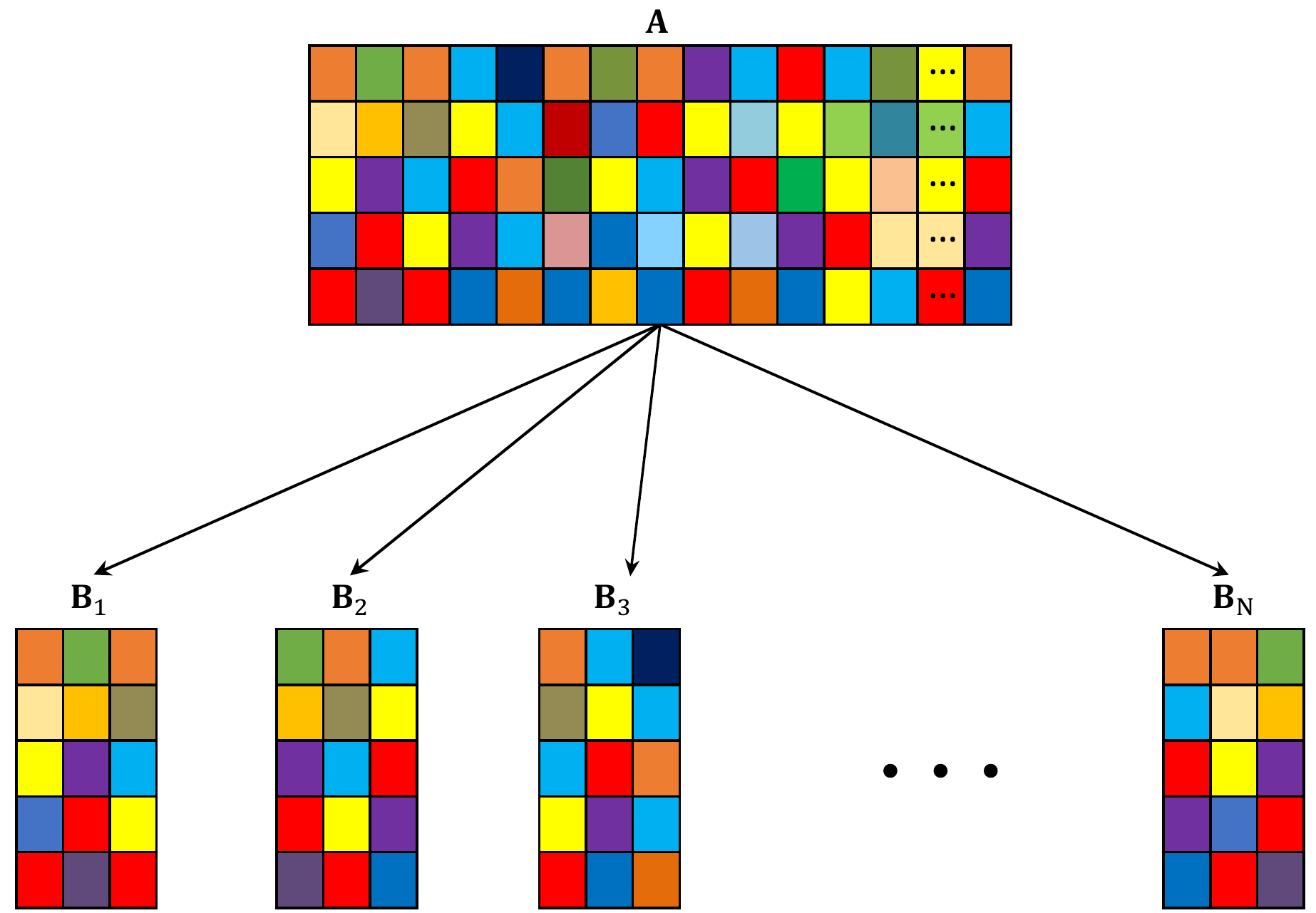}
	\caption{ Selection of sequence blocks from sensing matrix}
	\label{ss1}
\end{figure}

To distinguish the active nodes at the base station, each node must have a unique signature. In the proposed SB-CSMUD scheme, blocks of   spreading sequences are used as signatures.  A sensing matrix, $\mathbf{A} \in \mathbb{C}^{M\times N}$,  is generated by selecting random sequences from the unit circle such as $s_i(\nu) \sim \exp(2\pi\nu)$ with $\nu$ being a uniform distribution on the interval [0,1]. For each user a block of $D$ sequences are selected from the sensing matrix $\mathbf{A}$. The sequence block $\mathbf{B}_n$, $1\leq n \leq N$, is designed such that  the sequences at the same index of all sequence blocks are unique,  i.e., $\mathbf{B}_{i,d} \ne \mathbf{B}_{j,d},\:\:$ $ i, j \in \{1, 2, \cdots,  N \} , i\ne j$, where $ \mathbf{B}_{j,d}$ represents the $d$-th sequence of the $j$-th sequence block, $1 \le d \le D$.  While several procedures are possible for choosing the sequences for a sequence block, here for simplicity,  a sliding window based procedure is employed to select $D$ sequences from sensing matrix, $\mathbf{A}$. In Figure  \ref{ss1}, the design of sequence blocks for $D=3$ is shown. For user 1, the $\mathbf{B}_1$ consists of the first three sequences of the matrix, $\mathbf{A}$. Similarly $\mathbf{B}_2$ consists of second, third and fourth sequences and so on. \par 
\par 
\subsection{Processing at Sensor Node}
\begin{figure}[!h]
	\centering
	\includegraphics[scale=.47]{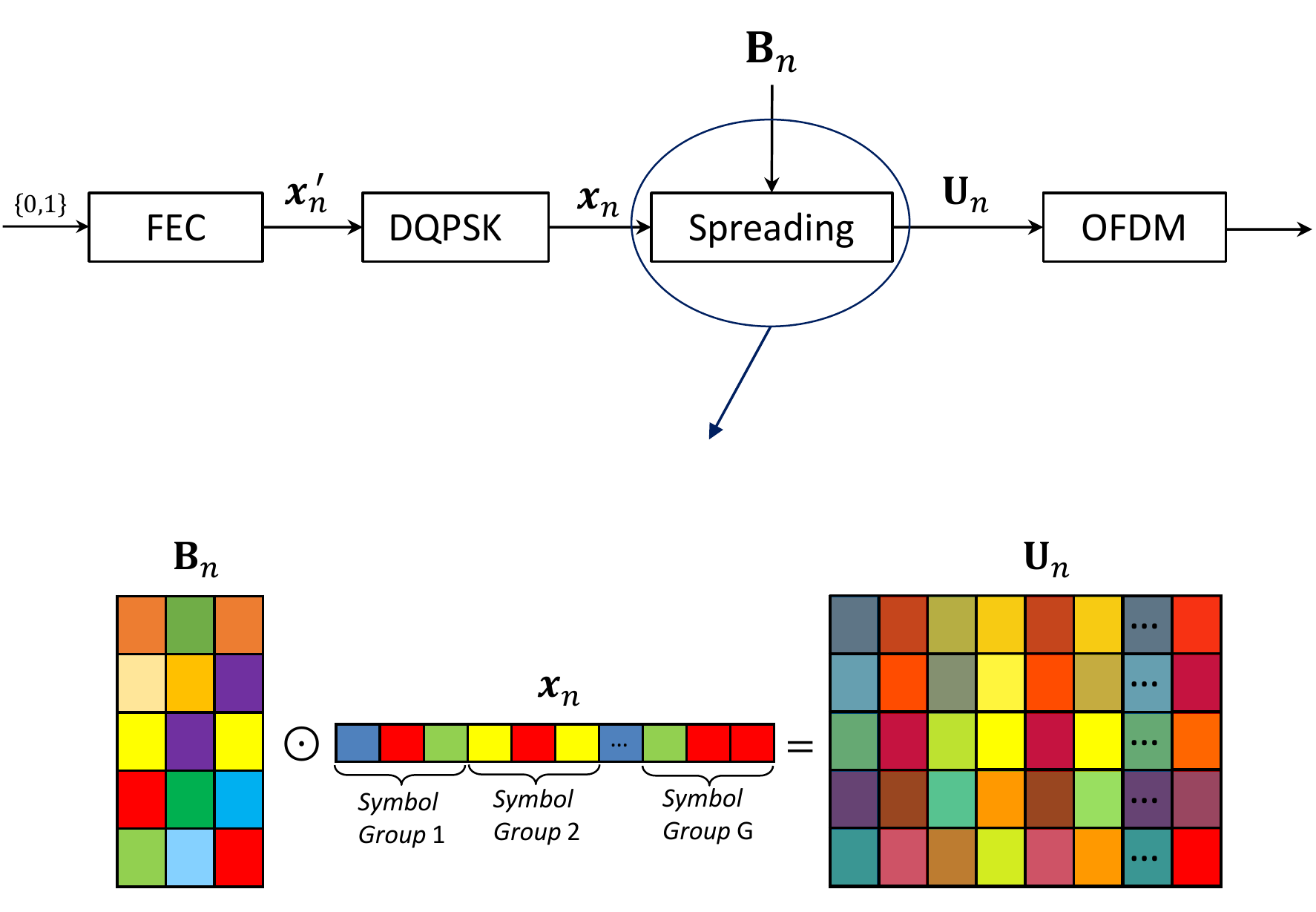}
	\caption{ Processing at node $n$: $\odot$ represents multiplication of first column $\mathbf{B}_n$ with fist symbol of a symbol group of  $\boldsymbol{x}_{_n}$, second column with second symbol  and so on }
	\label{spr}
\end{figure}

The data processing at a sensor node is depicted in Figure \ref{spr}. The active node transmits a data frame of $L_c$ bits which are selected from binary alphabet $\mathcal{A}$, whereas the inactive node is considered as a frame of $L_c$ zero bits.   The data of active nodes after channel coding are modulated  using DQPSK. DQPSK is used to facilitate the non-coherent detection at the receiver and therefore avoid channel estimation. The modulated data frame, $\boldsymbol{x}_{_n}\in \mathbb{C}^{1\times L}$, of user $n$ is divided into $G=L/D$ symbol groups, each symbol group having $D$ symbols. The symbols of each group are spread over the corresponding spreading sequences in the node specific sequence block.  An example for $D=3$ is shown in Figure \ref{spr}, where the first symbol of each group spreads over the first sequence in the sequence block, the second symbol over the second sequence and so on.\par
 The resultant spread signal frames of all the $N$ nodes   $\mathbf{U}_n \in \mathbb{C}^{M \times L}$, $1\leq n \leq N$,  are then transmitted using orthogonal frequency devision multiplexing.\par  
 
\subsection{Multiuser Detection}
The received signal after affected by the additive white Gaussian noise and channel fading is given in frequency domain as
\begin{equation}
\begin{split}
\mathbf{Y}   & =  \sum_{n=1}^{N} \diag{(\mathbf{h}_n)\mathbf{B}_n } \odot\mathbf{x}_n+\mathbf{W},\\
\end{split}
\end{equation}
where
\begin{equation}
\begin{split}
\mathbf{B}_n\odot \mathbf{x}_n  = & [\mathbf{B}_{n,1} \mathbf{x}_{n,1}, \mathbf{B}_{n,2}\mathbf{x}_{n,2}, \cdots,  \mathbf{B}_{n,D}\mathbf{x}_{n,D},\\
&  \mathbf{B}_{n,1}\mathbf{x}_{n,{D+1}}, \cdots, \mathbf{B}_{n,D}\mathbf{x}_{n,{2D}}, \cdots, \mathbf{B}_{n,D}\mathbf{x}_{n,{L}}],
\end{split}
\end{equation}
where $\mathbf{B}_{n,d}$ represents the $d$-th sequence of the $n$-th sequence block, $\mathbf{B}_{n,l}$ represents the the $l$-th symbol of the data frame of the $n$-th   node,   $\mathbf{W}\in \mathbb{C}^{M \times L}$ is the Guassian noise and   $\mathbf{h}_n \in \mathbb{C}^{M \times 1}$ consists of the user and subcarrier specific channel coefficients. The sequence block serves as a signature for the node at the receiver, where the activity is detected based on the average of the correlations between the sequence blocks and the received signal $\mathbf{Y}$. The activity detection step in Equation (\ref{ac1}) is redefined as 
\begin{equation}
I= \argmaxA_{1\leq j \leq N} \left(\Omega_j- \langle \mathbf{B}_{_j}, \mathbf{B}_ {(k)}\rangle \right),
\end{equation}
where $\mathbf{B}_ {(k)}$ is the detected sequence block in the previous iteration. The block correlation, $\langle \mathbf{B}_{_j}, \mathbf{B}_ {(k)}\rangle $ is given as 
\begin{equation}
\langle \mathbf{B}_{_j}, \mathbf{B}_ {(k)}\rangle  =   \frac{1}{D} \left( \sum\limits_{d=1}^{D} \frac {|\mathbf{B}_{j,d}^T   \mathbf{B}_{(k),d} |} {\| \mathbf{B}_{(k),d}\| \| \mathbf{B}_{j,d}\|  } \right),
\end{equation}
where $\mathbf{B}_{j,d}$ represents the $d$-th sequence of the $j$-th sequence block. The correlation contribution of the $j$-th block, $1\le j \leq N$, in the received signal is given as  
\begin{equation}
\label{c2}
\Omega_j= \frac{1}{L} \sum\limits_{\substack{l=1}}^{L}\sum\limits_{i=1}^{N} {\mathbf {x}}_i^{(l)} \cdot \langle\mathbf{B}_{_j},\mathbf{B}_{_i}\rangle+w_j^{(l)}.
\end{equation}
From Equation (\ref{c2}), it is clear that in the SB-CSMUD, the performance depends upon the block sequence instead of single sequence. The performance  determining factor of CSMUD is therefore redefined for the sequence block based GOMP as 
\begin{equation}
\label{mub}
\mu_B= \max\limits_{\tiny{\substack{i,j\\ i\neq j}}} \langle \mathbf{B}_{_i},\mathbf{B}_{_j} \rangle  \:\:\: i,j\in\{1,2,\hdots,N \}.
\end{equation}\par 
For the same random sensing matrix, $\mathbf{A}$,  the maximum correlation, $\mu$, in Equation (\ref{mu}), is between two sequences, whereas $\mu_B$ is the average of the correlations between blocks of $D$  sequences.  For the worst case, out of the $D$ correlations, $\mu_B$ can only have correlation of  a pair of sequences equal to $\mu$. The other $D-1$ correlations will be less than $\mu$. Therefore,   the  inequality,  $\mu_B < \mu$,    always exists, which ensures the improvement in the activity detection of the SB-CSMUD. 
The sequence block based GOMP algorithm for the multiuser detection is given in Algorithm \ref {alg:the_alg}. In line 1 of the algorithm, the activity is detected as the index of the sequence block that has maximum correlation with the residual $\mathbf{R}$ which is initialized as the received signal $\mathbf{Y}$.  The active nodes set $\Gamma$ is updated at each iteration with the newly detected node. For each symbol group, $g$, the data of the active nodes are estimated using least square estimation in line 2. In line 3, the residual is updated by subtracting the contribution of the estimated data from $\mathbf{Y}$. The algorithm stops when the number of iterations equals the number of active nodes or the energy of the residual becomes less than a predefined threshold, $\gamma$.    

\begin{algorithm}
	\SetKwInOut{Input}{Input}
	\SetKwInOut{Output}{Output}
	\SetKwInOut{Initialization}{Initialization}
	\SetKwInOut{Iteration}{Iteration}
	\label{alg:the_alg}
	\Input{$\mathbf{Y}, \mathbf{A}, G$}
	\Initialization {$q=0$, $\mathbf{R}^{0}=\mathbf{Y}, \Gamma=\emptyset$}
	\Iteration {$q \longleftarrow q+1$}	
	\SetNoFillComment
    \tcc{Activity Detection }	
			${I_q}\longleftarrow \argmaxA\limits_{ 1\leq n \leq N} \sum\limits_{g=1}^{G}\langle \mathbf{B}_n, \mathbf{R}_g^{q-1}   \rangle , $  \:\:\: 	$\Gamma^q = I_q \cup \Gamma^{q-1}$\\ 		
			\vspace{0.1cm}	
		    \tcc{Data Detection}
		    \vspace{0.1cm}
$\widehat{\mathbf{X}}_{{\Gamma^q,g}} = \mathbf{B}_{\Gamma^q}^\dagger \mathbf{Y}_g$   for  $1\leq g \leq G$                              \\
\vspace{0.1cm}
		    \tcc{Residual Update}
$\mathbf{R}_g^q =  \mathbf{Y}_g - \mathbf{B}_\Gamma \widehat{\mathbf{X}}_{{\Gamma^q,g}}$ \\
\vspace{0.1cm}
	\textbf{If} {$q= K$ or $\left \Vert \mathbf{R}^q\right \Vert < \gamma$}, stop \\
	\Output{$\widehat{\mathbf{X}}$ }  
	\caption{Sequence block GOMP}
\end{algorithm}

\section{Performance Analysis} \label{pa}
We analyzed the performance of the SB-CSMUD scheme in terms of DER and BER for various scenarios.  We assume that each active node transmits, $L_c= 100$ bits per data frame. An exponentially decaying channel with a path loss constant of two is assumed with block fading where the channel response remains same for 10 consecutive OFDM symbols.  The overloading factor is defined as $\lambda=N/M$. The  other simulation parameters are summarized in Table \ref{tb}. \par  
\begin{table}[]
	\centering
	\caption{Simulation parameters}
	\label{tb}
	\begin{tabular}{|l|l|}
		\hline
		Number of Nodes              & $N=40:20:120$                \\ \hline
		Length of spreading sequence & $M=20$                \\ \hline
		Activity probability         & $p_{_a}= 0.1:0.02:0.16$              \\ \hline
		Overloading factor           & $\lambda=2:6$      \\ \hline
		Channel coding               & Rate-1/2 Convolutional \\ \hline
		Modulation                   & DQPSK       \\ \hline
		Interleaver                  & Random              \\ \hline
		Delay spread length          & 1000 m               \\ \hline
		Fading model                 & Block fading        \\ \hline
	\end{tabular}
\end{table}\par
\begin{figure}[!h]
	\centering
	\includegraphics[scale=.6]{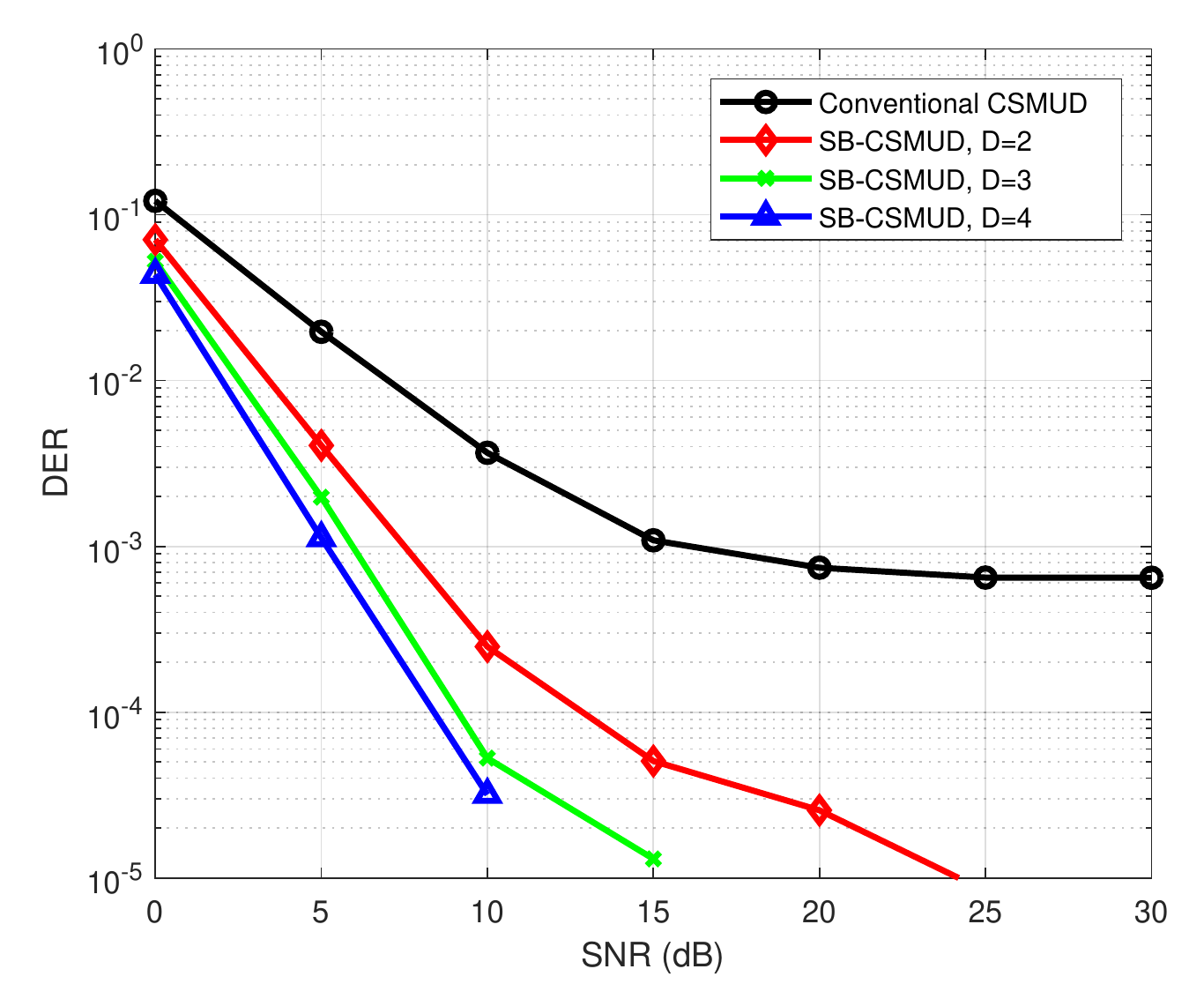}
	\caption{ Detection error rate for different sequence block size, $D$: $M=20, \: \lambda=3,\: \: p_a=0.1$ }
	\label{d1}
\end{figure}

  
The DER of the conventional and the SB-CSMUD schemes is compared in Figure \ref{d1}. The DER of the proposed SB-CSMUD scheme is significantly lower than that of the conventional scheme. It can be seen that at SNR = 10 dB, the DER of the conventional scheme is reduced by a magnitude of one using block size, $D=2$, $M=20$, $\lambda=3$ and $p_a=0.1$. Increasing the block size, $D$, further reduces the DER and it can be observed that at block size of four, a gain of magnitude two in DER is achieved at SNR = 10 dB.  However, at lower SNR, the reduction in DER with increasing $D$ is quadratic due to the fact that at lower SNR the noise significantly disrupt the signal. It can be observed from  Figure \ref{d1} that the  rate of reduction from $D=1$, i.e., the conventional case, to $D=2$ is much higher than that of the case from $D=2$ to $D=3$ and so on.\par  
\begin{figure}[!h]
	\centering
	\includegraphics[scale=.6]{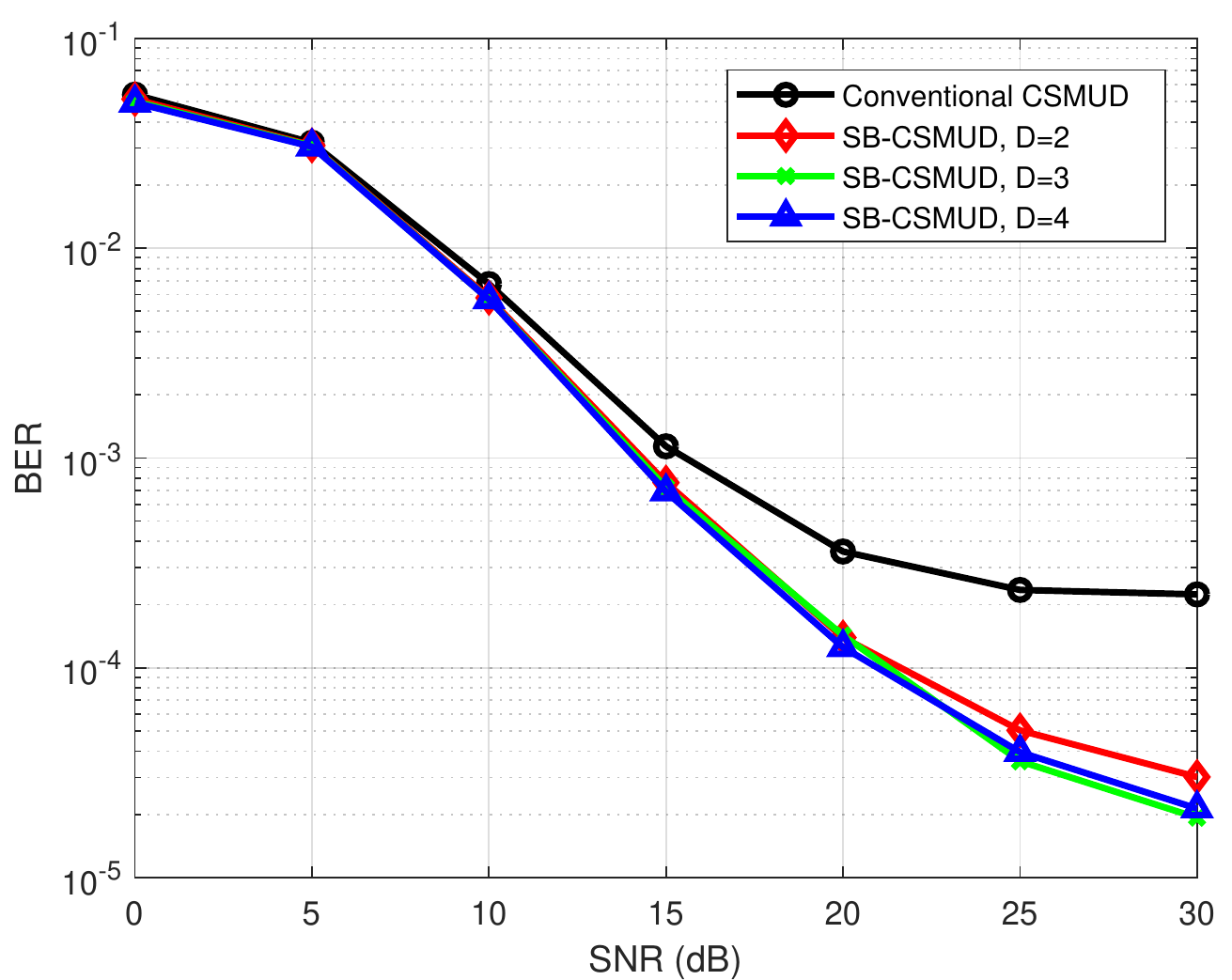}
	\caption{ Bit error rate for different sequence block size, $D$: $M=20, \: \lambda=3, p_a=0.1$  }
	\label{b1}
\end{figure}

In Figure \ref{b1}, it shows the corresponding BER comparison  of the schemes under the same settings of Figure \ref{d1}. The BER is significantly improved in the proposed SB-CSMUD scheme due to the accurate activity detection. We can find that the BER is reduced by a magnitude of one, at SNR = 30 dB. Referring to Figure \ref{d1}, it is evident that the reduction in BER comes from the reduction in the DER. The BER for $D \ge 2$ is nearly the same as that of  $D = 2$, the reason lies in that for   $D \ge 2$ the BER is dominated by the least square estimation errors. At lower SNR, although the activity is detected correctly, the improvement in BER of the conventional scheme  is not significant due to the higher number of  the least square estimation errors.\par 
\begin{figure}[!h]
	\centering
	\includegraphics[scale=.55]{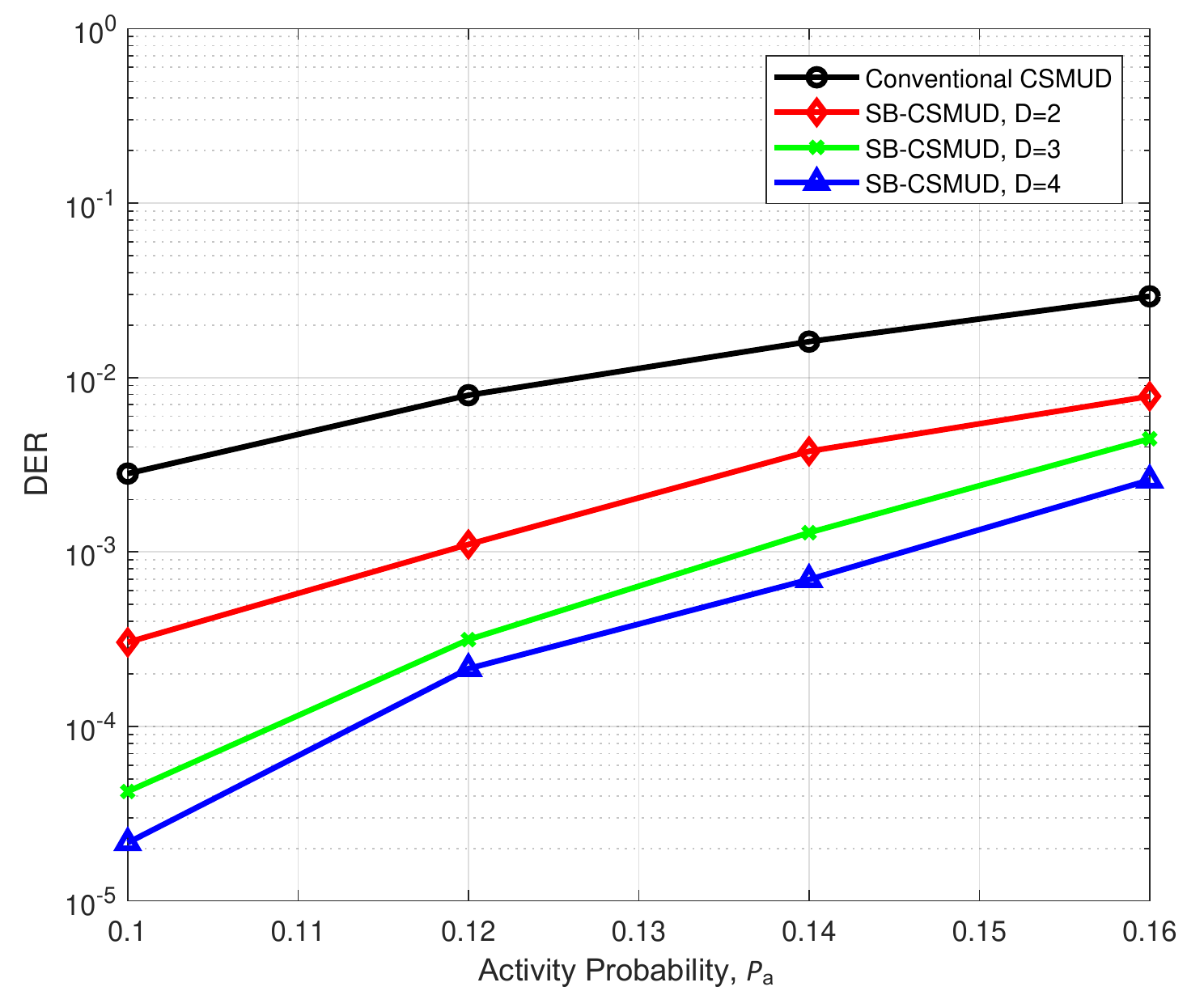}
	\caption{Effect of increasing the activity probability, $p_a$, on DER: $M=20, \: \lambda=3$, SNR = 10 dB}
	\label{p1}
\end{figure}

For a given SNR = 10 dB, it shows the effect of increasing the activity probability, $p_a$, on the DER in Figure \ref{p1}.  Increasing $p_a$, increases the number of active nodes, which increases the multiple access interference and ultimately leads to  activity detection errors. From Figure \ref{p1}, it is evident that the proposed scheme is more robust to multiple access interference and even at $p_a=0.16$, the DER is less than $10^{-2}$, whereas the DER of the conventional scheme is higher than $ 10^{-2}$ when  $p_a > 0.12$. Moreover, it can also be seen that increasing $D$ improves the performance, which shows that at higher, $p_a$,  the  DER can be further lowered by increasing $D$.  \par 
\begin{figure}[!h]
	\centering
	\includegraphics[scale=.6]{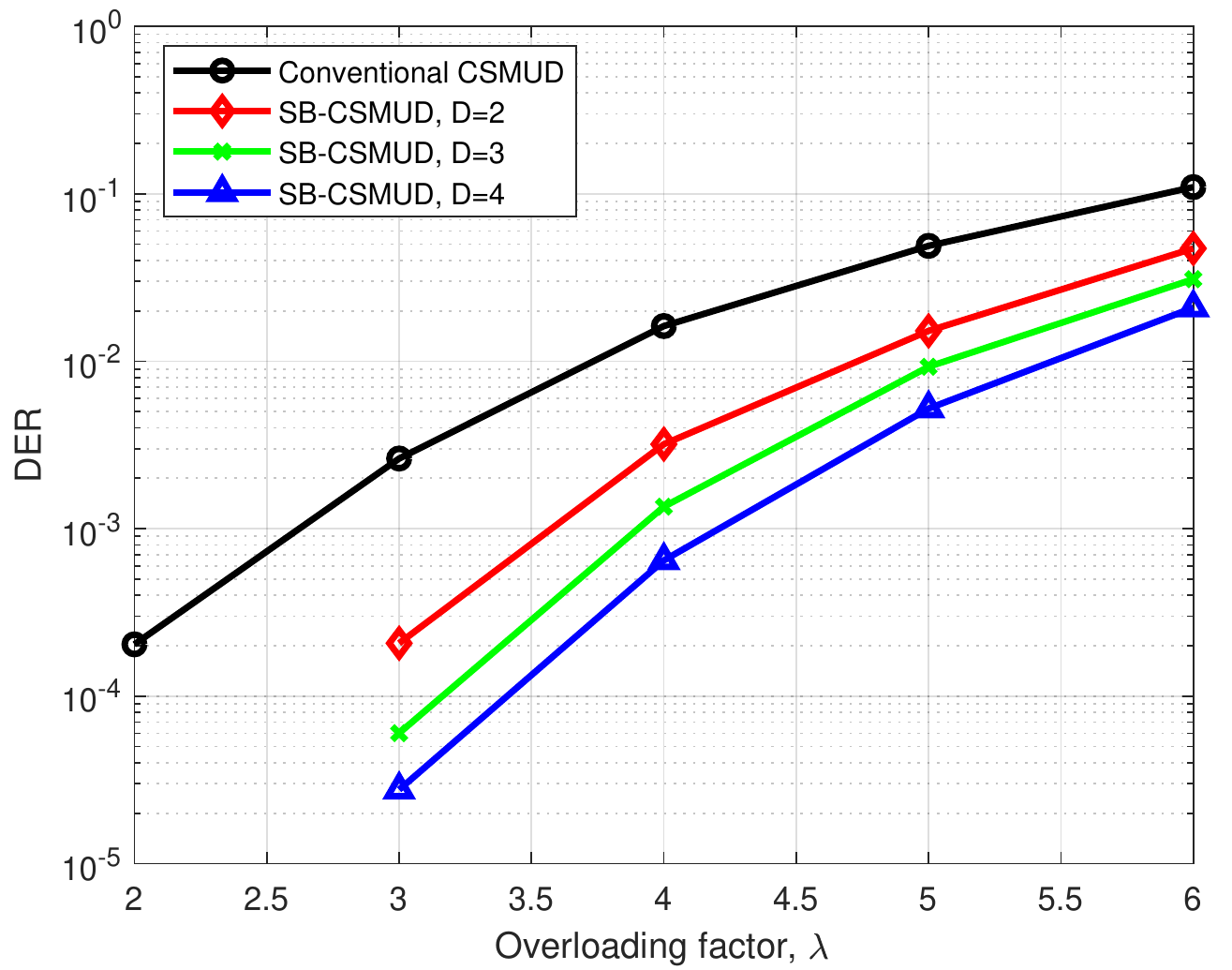}
	\caption{Effect of increasing the overloading factor, $\lambda$,  on DER: $ M=20, \:  p_a=0.1$, SNR = 10 dB}
	\label{k1}
\end{figure}

Figure \ref{k1} presents the effect of increasing the overloading factor, $\lambda$, on DER at SNR = 10 dB and $p_a=0.1$.  Increasing the overloading factor means increasing the number of spreading sequences for a fixed spreading factor. Therefore, for a fixed spreading factor, the maximum correlation increases with increase in $\lambda$, and  it becomes harder to correctly detect the active node. However, the increase in  correlation between the blocks is much smaller than that of the single sequence based conventional system.  Therefore, the DER of the SB-CSMUD  is far less than the conventional scheme. It can be observed from Figure  \ref{k1} that  DER is reduced by a magnitude of one at $\lambda=5$ for block size of $D=4$. Moreover, for $\lambda=2$, all of the nodes are correctly detected by the proposed scheme, whereas the conventional scheme has a DER of greater than $10^{-4}$. Furthermore, the DER at higher overloading also reduces with increasing the block size $D$. 

\section{Conclusion} \label {con} 
In this paper a sequence block based CSMUD  is proposed to improve the activity detection in a non-orthogonal MC-CDMA system for sporadic mMTC.  Accordingly, a sequence block based GOMP algorithm is designed to jointly detect the activity and data. The proposed scheme uses blocks of sequences as signatures for the users which  are generated from the predefined sensing matrix. As a result the performance is improved without reducing the spectral efficiency of the system.  The performance gain comes from the fact that the correlation between the sequence blocks is much lower than the correlation between single sequences. As compared to the conventional scheme, the proposed scheme reduces the DER by magnitude of two at SNR = 10 dB using a sequence block size of 4. The reduction in DER consequently improves the BER, specifically in the higher SNR region. Furthermore, it is shown that SB-CSMUD is more resilient at higher activity probability and in a highly overloaded system.In addition, the performance at these conditions can further be improved by increasing the block size.         
 \bibliographystyle{unsrt}
\bibliography{bibfile} 
  \end{document}